\documentstyle[a4,11pt]{article}
\def\slash{\!\!\!/}
\begin{document}     
\title{{\bf Heavy Quark Effective Fields as Operator-valued 
Distributions}
\vspace*{0.2in}
\thanks{Research partially supported by CICYT under contract AEN-96/1718}}
\author{Miguel Angel Sanchis-Lozano 
\thanks{E-mail: mas@evalvx.ific.uv.es ; 
Phone: +34 6 386 4752 ; 
Fax: +34 6 386 4583} \\
\\
\it Departamento de F\'{\i}sica Te\'orica and IFIC\\
\it Centro Mixto Universidad de Valencia-CSIC\\
\it 46100 Burjassot, Valencia, Spain}
\maketitle 
\vspace*{1.0in}
\abstract{We look at effective fields defined in the heavy-quark 
effective theory as operator-valued generalized functions on Minkowski 
space-time to be averaged with physically suitable smoothing functions 
(Gaussians of typical width the hadronic size) leading to operator-valued
distributions in Hilbert space. One-heavy-quark states are 
thus represented by normalizable wave packets displaying a particle-like
behaviour at the characteristic hadronic time scale ${\Lambda}_{QCD}^{-1}$. 
We examine some consequences relative to the average kinetic energy of 
heavy quarks in hadrons to avoid inconsintences within this formalism.}
\large
\vspace{-15.5cm}
\begin{flushright}
  FTUV/97-43 \\
  IFIC/97-73
\end{flushright} 
\vspace{15.cm}
\begin{small}
\end{small}
\newpage
\section{Introduction}
In recent years the physics of hadrons containing a heavy quark
has been analyzed by means of an effective theory for
the strong interaction (HQET) based on an expansion in the inverse
heavy quark mass \cite{neub}. This approach shows the 
existence of new symmetries at leading order, dealing systematically
with the corrections arising from higher order terms, and allowing
numerous fruitful phenomenological applications among which the extraction
of the Cabibbo-Kobayashi-Maskawa matrix element 
${\mid}V_{cb}{\mid}$. Nevertheless, such a powerful framework still 
suffers from some internal ambiguities. Specifically, typical parameters
like the average kinetic energy of the heavy quark in a hadron 
(needed for a precise determination of ${\mid}V_{cb}{\mid}$ from
$B$ decays) are under intense discussion, and whose precise physical
definiteness has been questioned (see \cite{bigi,neu1}
and references therein). 
\par
In constructing HQET a cornerstone is the observation that
massive quarks in heavy-light hadrons exchange momenta with
the light degrees of freedom typically of order
${\Lambda}_{QCD}$, the characteristic scale of the strong
interaction, much smaller than the heavy quark mass $m_Q$. 
Therefore and much contrary to light quarks, a heavy quark should propagate
along a particle-like trajectory in space - with an almost constant 
velocity - until the decay of the hadron takes place. This physical 
motivation (with vivid analogies in early literature for the heavy quark
such as baseball \cite{isgur} or even cannonball \cite{geor}) shows that HQET
resembles more a classical approximation (e.g. no heavy-quark pair
production is allowed) than a non-relativistic
one, at least as far as the long-distance aspects of the theory are
concerned. 
\par
These remarks suggest the split of the
heavy quark four-momentum into two pieces:
\begin{equation}
p_Q\ =\ m_Qv\ +\ k_Q
\end{equation}
where the first term in the rhs represents the large mechanical part as $v$  
is the hadron's velocity (although this is not necessary because of the
so-called reparametrization invariance of the theory \cite{luke}) 
and $k_Q$ denotes the $\lq\lq$residual momentum" arising from the
predominantly soft interaction of the heavy quark with gluons.
\par
Squaring $p_Q$ we shall write
\begin{equation}
p_Q^2\ =\ m_Q^2\ +\ {\Delta}
\end{equation}
with ${\Delta}$ denoting a measure of the off-shellness of the
heavy quark, i.e.
\[
{\Delta}\ =\ p_Q^2-m_Q^2\ =\ 2m_Qv{\cdot}k_Q\ +\ k_Q^2
\]
where in the limit $m_Q{\rightarrow}\infty$, 
${\Delta}/m_Q^2{\rightarrow}0$.
\par
As a simple but illustrative exercise, let us suppose at a given
instant a heavy quark on-shell (with a particular choice for $m_Q$) 
and at rest in the hadron rest frame, absorbing a soft on-shell gluon 
whose four-momentum is $({\Lambda}_{QCD},\vec{{\Lambda}}_{QCD})$. Looking 
at the interaction as a real process, energy-momentum 
conservation implies that after the absorption the heavy quark mass 
$m_Q^*$ would shift according to
\[
m_Q^{*2}\ =\ m_Q^2\ +\ 2m_Q{\Lambda}_{QCD}
\]
that is
\[
m_Q^*\ {\simeq}\ m_Q\ +\ {\Lambda}_{QCD}
\] 
while its kinetic energy should be ${\simeq}\ {\Lambda}_{QCD}^2/2m_Q$. 
Notice, however, that the absorbed energy by the heavy quark actually is
${\Lambda}_{QCD}$. (A similar reasoning applies to the emission of a 
soft gluon.)
\par
According to the energy-time uncertainty principle it is possible
to interpret that during a characteristic time interval ${\Delta}t$ given by
\footnote{We keep all numerical factors - they cancel out at the end of the 
calculation anyway  - although, of course, only the order of magnitude is 
significant. Factor 3 is associated with the fact that we are
considering three-dimensional motion}
\begin{equation}
{\Delta}t\ {\simeq}\ \frac{3}{2}\ {\Lambda}_{QCD}^{-1}
\end{equation}
the heavy quark propagates virtually free over a distance of at
least its Compton wavelength, after which its
(total) energy is changed by an amount ${\simeq}\ {\Lambda}_{QCD}$, 
varying both its mass and kinetic energy accordingly. 
In fact, this assumption is on the grounds for the use of the
static approximation in computing the quark-antiquark
(QCD-inspired) potential for heavy quarkonium, where 
${\Lambda}_{QCD}^{-1}\ {\simeq}\ 10^{-23}$s is the typical time 
required by the gluonic fields to adjust themselves to the movement of 
the heavy quarks \cite{lepage}. Furthermore, ${\Lambda}_{QCD}^{-1}$ is 
as well the characteristic time scale for building up heavy-light or 
heavy-heavy hadrons subsequently to a heavy-flavour production process 
\cite{bigi1}. 
\par
As a consequence of many successive soft interactions between the heavy 
quark and the light constituents the former acquires a mean kinetic energy 
while its mass $\lq\lq$fluctuates" about $m_Q$ with typical deviation
${\Lambda}_{QCD}$ (that is 
${\mid}{\Delta}{\mid}\ {\simeq}\ 2m_Q{\Lambda}_{QCD}$). Certainly this
is an idealization since we have ignored the existence of short-distance
physics. Nonetheless, hard gluons play little role in the structure
of low-lying hadronic states and, in particular, the HQET operator
of the non-relativistic kinetic energy is not multiplicatively renormalized
as a consequence of the reparametrization invariance already 
mentioned \cite{luke}. We shall turn to this point in Section 3, however.
\par
All the above considerations suggest writing a plane-wave Fourier 
expansion for the fermionic field $Q_v{(x)}$ (positive frequencies)
\cite{mas,korner} corresponding to an almost on-shell heavy quark 
\footnote{Heavy anti-quarks would be treated in the same way. We also
omit any reference to colour and flavour} moving inside
a hadron with four-velocity $v$   
\begin{equation}
Q_v(x)\  =\  \int d{\mu}(p)\ \sum_{r}\ 
b_r(\vec{p})\ u_r(\vec{p})\ e^{-ip{\cdot}x}
\end{equation}
where we have required that all the Fourier components satisfy
the same off-shellness condition imposed by Eq. (2), that is
\begin{equation}
p^2\ =\ m_Q^2\ +\ {\Delta}
\end{equation}
\par 
Hence $d{\mu}(p)$ stands for the invariant measure although
$p$ does not exactly lay on the upper mass-shell hyperboloid 
(corresponding to the $\lq\lq$assumed" $m_Q$ value)
since $p^0=\sqrt{m_Q^2+{\Delta}+\vec{p}^2}$. On the
other hand, $b_r(\vec{p})$ is the usual 
annihilation particle operator, for a heavy quark in this case.
\par
Let us redefine the momenta of the Fourier components in (4)
according to HQET as 
\begin{equation}
p\ =\ m_Qv\ +\ k
\end{equation}
where $k$ represents the Fourier residual four-momentum. Let us
stress that in our notation the latter must not be confused with the 
residual four-momentum $k_Q$ characterizing a particular hadronic state as 
an expectation value (see Ref. \cite{mas}).
\par
On the other hand, velocity-dependent effective fields in HQET are 
introduced as \cite{neub}: 
\begin{equation}
h_v(x)\ =\ e^{im_Qv{\cdot}x}\ P_{+}\ Q_v(x)       
\end{equation}
\begin{equation}
H_v(x)\ =\ e^{im_Qv{\cdot}x}\ P_{-}\ Q_v(x)
\end{equation}
where the projector operators $P_{\pm}=(1{\pm}v{\slash})/2$ leave
only the upper (lower) components in the hadron rest frame. 
As is well-known, the $\lq\lq$small" components $H_v$ can be
removed by means of the (classical) field equations yielding
the following long-distance Lagrangian at order $1/m_Q$ expressed in
standard notation as
\begin{equation}
L_{eff}\ =\ \overline{h}_viv{\cdot}Dh_v\ +\ 
\frac{1}{2m_Q}\overline{h}_v(iD_{\bot})^2h_v\ +\ 
\frac{g}{4m_Q}\overline{h}_v{\sigma}_{\alpha\beta}G^{\alpha\beta}h_v
\end{equation}
where the second term in the rhs stands for the kinetic energy
arising from the off-shell residual motion of the heavy quark.
The third term describes the chromomagnetic interaction of the heavy quark 
with the gluon field \cite{neub}.
\par
As far as the heavy quark could be considered (quasi)-free during
the tiny time interval (3), from comparison between (4) and (7) 
the $h_v$ field can be written as
\begin{equation}
h_v(x)\ =\ \frac{1+v{\slash}}{2}\ \int\ d{\mu}(k)\ 
\sum_{r}\ b_r(\vec{v},\vec{k})\ u_r(\vec{k})\ e^{-ik{\cdot}x}
\end{equation}              
where $b_r^{\alpha}(\vec{v},\vec{k})$ denotes 
the annihilation operator for a heavy quark
with residual spatial momentum $\vec{k}$ in a hadron moving with 
four-velocity $v$; $d{\mu}(k)$ stands again for the invariant measure 
now expressed in terms of the Fourier residual momentum. (The zero component 
of the residual four-momentum $k$ in the hadron rest frame must satisfy:
$k^0\ =\ \sqrt{m_Q^2+\vec{k}^2+{\Delta}}\ - m_Q$ corresponding to the 
small off-shellness condition derived from (5) 
$k^2+2m_Qv{\cdot}k={\Delta}$, for positive frequencies.)
\par
Observe that, as written in Eq. (10), 
the $\overline{h}_v(x)$ field creates a heavy quark at point $x$ 
of space-time as a superposition of single-particle states with momenta 
ranging over the entire $\vec{k}$ domain. Nevertheless, one may
introduce an ultraviolet cut-off eliminating those
components with residual momenta greater than the heavy quark
mass $m_Q$. This corresponds to ignore those Fourier components
whose wavelengths are smaller than the Compton 
wavelength of the heavy quark, amounting to a position
uncertainty of this order as is well-known from Quantum Theory
\footnote{In one-particle relativistic theory this amounts to a
zigzag motion traditionally known as zitterbewegung \cite{lurie}. This 
oscillatory motion, involving negative-energy states, present even in the 
hypothetical case of a {\em free} quark, should
be distinguished from the residual motion due to the soft interaction
with the light constituents of the hadron}. 
\par
Moreover, in constructing the low energy effective theory for heavy quarks 
one has still to add a further restriction to the Fourier expansion of the
fields, limiting the range of the  $\vec{k}$ components to values of the 
order of ${\Lambda}_{QCD}$. Note that this condition amounts to 
a new constraint not completely equivalent to the small virtuality 
already imposed by means of Eq. (5). Indeed, the smallness of 
$\vec{k}$ implies the almost on-shellness condition but the converse is
not necessarily true. In fact, removing those components with large 
residual spatial momentum in the Fourier expansion of the fields is 
similar to integrating away the high-velocity states in the functional
integral formulation of Ref. \cite{italia}.  
\par
Short-distance effects involving large virtual momenta
\footnote{As stressed in \cite{bigi}
the normalization scale ${\mu}$ should lie somewhere in the range
${\Lambda}_{QCD}<<{\mu}<<m_Q$ in the context of the Wilsonian
(operator product expansion) approach to the heavy-quark theory. 
In practice ${\mu}\ {\simeq}$ several units ${\times}$ ${\Lambda}_{QCD}$}
can be included in the effective theory in a perturbative way using 
renormalization group techniques in a procedure called matching \cite{neub}.

\section{Effective Fields as Operator-valued Distributions}
In HQET, hadron states are usually identified with the eigenstates
of the leading term in the Lagrangian (9) (corresponding to
$m_Q{\rightarrow}\infty$) supplemented with
the standard QCD Lagrangian for the light quarks and gluons \cite{neub}. 
However, as emphasized in \cite{grin2}, heavy-quark states in Hilbert space 
should be associated (to some arbitrariness) with a certain 
velocity $v$ {\em and} residual momentum $k$. Single-particle (Fock) 
states are thus obtained by letting the creation operators
$b_r^+(\vec{v},\vec{k})$ act on the vacuum. In the infinite mass limit 
(or in practice for $m_Q>>{\Lambda}_{QCD}$) any one-particle state
with a given value of $p_Q=m_Qv+k$ could be used to represent the
state ${\mid}\ p>$ of the full theory. In this work, however, we shall
consider a superposition of states (wave packet) for describing 
a heavy quark localized in a definite region of space. In fact, 
there is a framework to implement in a natural way the Fourier
content of the heavy quark effective fields based on the
concept of operator-valued distributions.
\par
 
Since the beginning of quantum field theory, it was pointed out by
Heisenberg that the measurement of a field quantity 
at a space-time point (though itself a useful mathematical entity) must
be impossible (see Ref. \cite{wight} for an interesting historical 
account of quantum field theory). Indeed, in experiments the field strength 
is always measured not at a mathematical point. 
In this respect Bohr and Rosenfeld \cite{bohr} early argued 
that the real meaning of fields was related to
average values in space-time regions. 
\par
To become properly defined operators in Hilbert space, fields 
($\Phi$) have to be smoothed by suitable functions of 
sufficiently regular behaviour on Minkowski space:
\begin{equation}
\Phi(f)\ =\ \int\ d^4x\ \Phi(x)\ f(x)
\end{equation}  
where $f$ belongs to a $\lq\lq$test" function space, usually
taken as the space $S$ of infinitely often differentiable
functions decreasing as well as their derivatives faster than any
power as $x$ moves to infinity \cite{bogo}. The resulting functional 
\cite{schwartz} is an operator-valued (tempered) distribution, which acting 
on the vacuum of Hilbert space of states generates normalizable 
one-particle states or wave packets.
\par  
On the other hand, one usually has to deal with several fields, each
of which may have tensor or spinor components. Thus Eq. (11) must
be generalized as

\begin{equation}
\Phi(f)\ =\ \sum_{r=1}^{n}\ \int\ \Phi_{r\alpha}(x)\ f_r^{\alpha}(x)\ d^4x
\end{equation}
where the index $r$ denotes the type of field and $\alpha$
stands for each component.
\par
In this paper our intention is not to present a rigorous development of
HQET based on the framework of the {\em general} quantum field theory 
\cite{bogo}. Rather we shall make use of some ideas underlying 
the concept of smeared fields to gain insight into several 
aspects of the effective theory for heavy quarks in hadrons. Indeed, working
in the hadron rest frame and due to the projectors 
$P_{\pm}$ only the two upper components of the spinor field are not null. 
Thus we shall dispense with the indices $r$ and
$\alpha$ of the the smoothing function which becomes
just a c-number. (In general, test functions should display a tensor or
spinor character in order to fulfil the Wightman axiom relative to
the covariance transformation properties of relativistic quantum fields 
\cite{bogo}.)
\par
Since (heavy) quarks are confined in the hadron $\lq\lq$volume", 
it seems natural to introduce a Gaussian (peaked at the 
center of the hadron) as a space averaging function at a given
instant chosen as $t=0$
\begin{equation}
f(x)\ =\ \frac{1}{(2\pi)^{3/4}{\sigma}_0^{3/2}}
\exp{\biggl[-\frac{{\vec{x}^2}}{4{\sigma}_0^2}\biggr]}\ \delta(t)
\end{equation}
where ${\sigma}_0$ should be interpreted as a measure of the
the typical hadronic size, i.e. the hadronic radius ${\simeq}\ 1$ fm.
\par
In (residual) momentum space, the Fourier transform reads 
\begin{equation}
\tilde{f}(\vec{k})\ =\ \frac{1}{(2\pi)^{3/4}\tilde{\sigma}^{3/2}}
\exp{\biggl[-\frac{\vec{k}^2}{4\tilde{\sigma}^2}\biggr]}
\end{equation}
where ${\sigma}_0=1/2\tilde{\sigma}$.
Therefore, the smeared out annihilation operators are defined as
\begin{equation}
b_r(\tilde{f})\ =\ \int\ d\mu(k)\ \ \tilde{f}(\vec{k})\ 
{\varphi}_r(\vec{k})\ b_r(v,\vec{k})
\end{equation}
with ${\varphi}_r$ standing for the upper components
of the spinor. Then it follows that the smeared heavy-quark 
effective field can be written as
\begin{equation}
h_v(f)\ =\ N\sum_{r=-1/2}^{1/2}\ b_r(\tilde{f})
\end{equation}
where $N$ is the appropriate normalization factor in HQET \vspace{0.1in} 
\cite{korner}.
\par
\section{Heavy-Quark States in Hilbert Space}
When acting on the vacuum, $\overline{h}_v(f)$ generates a normalizable
state as an ensemble of plane-wave states with momenta about $m_Qv$, 
corresponding to a minimum wave packet with the choice (14).
Let us remark that the velocity superselection rule of HQET 
\cite{geor} - which forbids combinations from different sectors of 
Hilbert space related to different $v$'s - does not actually apply in
this case since the residual momenta correspond to quantum
fluctuations around $m_Qv$. 
\par
In order to study the evolution of the wave packet in space-time
according to the Schr\"{o}dinger picture, let us write at $t=0$
\begin{equation}
<x{\mid}\overline{h}_v(f){\mid}0>\ =\ N\ \sum_{r=-1/2}^{1/2}\ 
\int\ d\mu(k)\ \tilde{f}(\vec{k})\ {\varphi}_r(\vec{k})\ e^{-i\vec{k}\vec{x}}
\end{equation}
where the Gaussian function $\tilde{f}$ actually restricts the
integral to small values of $\vec{k}$. 
\par
The $\tilde{\sigma}$ parameter in Eq. (14) can be interpreted as the 
root-mean-square residual momentum (for each spatial component) of 
the heavy quark inside the hadron,
\begin{equation}
<\vec{k}^2>\ =\ 3\ \tilde{\sigma}^2
\end{equation}
which, in turn, can be related to the average non-relativistic kinetic energy 
$K$ 
\begin{equation}
K\ =\ \frac{<k^2>}{2m_Q}\ =\ 
\frac{3\ \tilde{\sigma}^2}{2m_Q}
\end{equation}
\par
On the other hand, the expectation value of the
$1/m_Q$ kinetic term of the HQET Lagrangian  in (9)
is usually written as 
\begin{equation}
K^{HQET}\ =\ 
\frac{-<H(v){\mid}\overline{h}_v(iD_{\bot})^2h_v{\mid}H(v)>}
{2m_Q}\ =\ \frac{-{\lambda}_1}{2m_Q}
\end{equation}
with a mass-independent normalization of states. The ${\lambda}_1$ 
parameter together with $\overline{\Lambda}$ and ${\lambda}_2$ \cite{neub} 
(related to the light degrees of freedom and the chromomagnetic
interaction respectively) are basic quantities of HQET, playing a 
fundamental role in many of its phenomenological applications.
\par
Although the HQET kinetic operator is not multiplicatively renormalized, 
its mixing with the identity operator under ultraviolet renormalization
should lead to an additive quadratically divergent contribution to 
${\lambda}_1$ which has to be non-perturbatively subtracted \cite{maiani}
in the quest for a $\lq\lq$physical" quantity 
${\lambda}_1^{phys}$ \cite{neu1,gimenez}. Although the use of a 
hard ultraviolet cut-off delivers
matrix elements from renormalon ambiguities, unfortunately, 
the arbitrariness in the choice of the subtraction scheme reintroduces 
the ambiguity. Such an uncertainty, however, is expected to be small
if a Lorentz-invariant regularization is employed since then the mixing
occurs at next-to-leading order (i.e. two-loop order), as explicitly 
shown in \cite{neu1}. On 
the other hand, determinations of ${\lambda}_1$ using dimensional 
regularization are contaminated by the renormalon ambiguity problem.
\par
Hence, in our characterization of heavy-quark states by wave packets 
we shall assume $<\vec{k}^2>=-{\lambda}_1^{phys}$, where current numerical
values of the last parameter in literature may be possibly affected by 
ambiguities of order ${\Lambda}_{QCD}^2$.
\par

As already mentioned, the action when the field operator 
$\overline{h}_v(f)$ acts on the vacuum is to create a wave packet 
of typical width ${\sigma}_0$ in ordinary space at
$t=0$. However, as time elapses the wave packet representing the 
$\lq\lq$temporarily free" heavy-quark spreads out, its spatial width becoming 
increasingly larger according to \cite{bohm}
\begin{equation}
\sigma(t)\ =\ {\sigma}_0\ 
\biggl(1\ +\ \frac{4\tilde{\sigma}^4t^2}{m_Q^2}\biggr)^{1/2}
\end{equation}
whereas the momentum width $\tilde{\sigma}$ remains the same. 
In order for the wave packet description of localized 
(within the hadron volume) massive quarks to remain meaningful, one 
must require that
\begin{equation}
\tilde{\sigma}^2\ <<\ \frac{m_Q}{2\ t}
\end{equation}

The characteristic time $t$ of the above expression can be
identified with ${\Delta}t$ given by Eq. (3) \footnote{Tentatively
identifying $t$ with the typical lifetime of B mesons, of the order of 
the psec ($10^{-12}$s) makes no sense at all
since in order to satisfy the inequality (22) the initial wave packet 
should be too narrow in momentum space, or equivalently exceedingly
large in ordinary space in contradiction with the typical size of a 
hadron where the heavy quark is to be confined. Therefore it seems natural 
to consider the hadron lifetime divided in tiny time intervals
($\simeq{\Lambda}_{QCD}^{-1}$ in the hadron rest frame) during which 
the free wave packet description is meaningful}.  
Therefore setting $t\ {\simeq}\ (3/2)\ {\Lambda}_{QCD}^{-1}$ 
in Eq. (22) we get the condition:
\begin{equation}
\tilde{\sigma}^2\ <<\ \frac{1}{3}\ m_Q{\Lambda}_{QCD}
\end{equation}
and with the aid of (18), one can deduce that \footnote{Any other physical
scale of the order ${\Lambda}_{QCD}$ such as $\overline{\Lambda}$ could be 
used instead} 
\begin{equation}
-{\lambda}_1^{phys}\ <<\  m_Q\ {\Lambda}_{QCD}
\end{equation}
in spite of ambiguities of order ${\Lambda}_{QCD}^2$. As a matter of fact, 
what the above strong inequality actually 
means is that, for consistency, $-{\lambda}_1^{phys}$ must 
be much smaller than the scale of the light degrees of freedom times
the heavy quark mass, both sides being ambiguous to a similar extent which
should not invalidate it though.
\par
On the other hand, the uncertainty principle also implies 
a lower bound to the physical value of the heavy quark residual 
square momentum. Hence we conclude that the following double inequality 
must be satisfied for a particle-like description of heavy-quarks:
\par
\begin{equation}
{\Lambda}_{QCD}^2\ {\leq}\ -{\lambda}_1^{phys}\ <<\ m_Q\ {\Lambda}_{QCD}
\end{equation}
\par
Observe that the rhs of expression (25) crudely states that the
kinetic energy of the heavy quark must be much smaller than
${\Lambda}_{QCD}$, i.e. much smaller than the energy exchange
with the gluonic cloud. (In the infinite mass limit the massive
quark would become a pure static source of colour.)
In particular, setting standard numerical values for the $b$-quark mass 
and ${\Lambda}_{QCD}$ we determine the following 
range (in GeV$^2$ units):
\begin{equation}
0.06\ {\leq}\ -{\lambda}_1^{phys}\ <<\ 1
\end{equation}
\par
Consequently note that the last expression favours relatively small
values for the $-{\lambda}_1^{phys}$ parameter, as for example
the result quoted by Neubert in \cite{neu3} making use of the 
field-theory analog of the virial theorem
for heavy-light hadrons \cite{neu4} (see also Ref. \cite{gremm} for
phenomenological determinations from $B$ and $D$ meson decays). 
Nevertheless, since our expression (26) is in reality an order-of-magnitude 
estimate, somewhat larger values for the average square momentum 
\cite{bigi,fazio} do not necessarily 
lead to contradiction with a particle-like behaviour
of the heavy \vspace{0.1in} quark.
\par
What about charm quarks in \vspace{0.1in} hadrons? 
\par
Setting $m_c=1.5$ GeV and ${\Lambda}_{QCD}=0.25$ GeV
the double inequality (in GeV$^2$ units)
\begin{equation}
0.06\ {\leq}\ -{\lambda}_1^{phys}\ <<\ 0.4 
\end{equation}
now leaves little room for the meaning of a localized $c$-quark 
in a hadron. Needless to say, if light quarks are considered 
($m_q\ {\simeq}\ {\Lambda}_{QCD}$) it becomes apparent that the 
inequalities (25) can not be simultaneously satisfied any more and the 
formalism is not appropriate at all. In the language of ordinary Quantum 
Mechanics light quarks would display a wave-like character.
\par
\section{Conclusions}
In this work we have averaged effective fields $h_v(x)$ defined in HQET 
for almost on-shell heavy quarks by means of smearing functions (Gaussians
spatially peaked at the center of the hadron) yielding operator-valued
distributions in Hilbert space. Thus, smoothed field operators acting on 
the vacuum create normalized wave packets representing heavy quarks
as superposition of plane-wave 
states with residual momenta around $m_Qv$. Their subsequent time 
evolution in ordinary space leads, however, to increasingly larger 
spatial spreads, becoming incompatible with a particle-like
interpretation for a typical $B$ meson lifetime.  In fact, the 
validity of this description has to be limited to the characteristic
hadronic time scale ${\Lambda}_{QCD}^{-1}$. Thereby, we conclude
that the double inequality (25) heuristically suggests
relatively small absolute values of the HQET parameter ${\lambda}_1^{phys}$
for a $b$-quark in a hadron, either meson or baryon. 
Such a picture becomes doubtful for $c$-quarks and clearly
meaningless for light \vspace{0.1in} quarks.
\par
\end{document}